\documentclass[useAMS,usenatbib]{mn2e}

\usepackage{amsmath}

\usepackage{epsfig}
\usepackage{epstopdf}
\usepackage{lscape} 
\usepackage{natbib}
\usepackage{tabularx}
\usepackage{multirow}
\usepackage{amssymb}
\usepackage{gensymb}
\usepackage{color}

\def\gtrsim{\mathrel{\hbox{\rlap{\hbox{\lower4pt\hbox{$\sim$}}}\hbox{$>$}}}}

\title{Constraints on the FRB rate at 700-900 MHz}
\author[Connor et al.]{Liam Connor$^{1,2,3}$\thanks{E-mail:\ connor@astro.utoronto.ca},
Hsiu-Hsien Lin$^{4}$,
Kiyoshi Masui$^{5}$,
Niels Oppermann$^{1}$,
\newauthor \,Ue-Li Pen$^{1,3,6,7}$,
Jeffrey B. Peterson$^{4}$,
Alexander Roman$^{4}$,
Jonathan Sievers$^{8}$, 
\\
$^1$ Canadian Institute for Theoretical Astrophysics, University of Toronto, M5S 3H8 Ontario, Canada
\\
$^2$ Department of Astronomy and Astrophysics, University of Toronto, 
M5S 3H4 Ontario, Canada
\\
$^3$ Dunlap Institute for Astronomy and Astrophysics, University of Toronto,
Toronto, ON M5S 3H4, Canada
\\
$^4$ McWilliams Center for Cosmology, Carnegie Mellon University, Department of Physics, 5000
Forbes Ave, Pittsburgh, PA, 15213, USA
\\
$^5$ Department of Physics and Astronomy, University of British Columbia, 6224 Agricultural Rd, Vancouver, BC, V6T 1Z1, Canada
\\
$^6$ Canadian Institute for Advanced Research, Program in Cosmology
and Gravitation
\\
$^7$ Perimeter Institute for Theoretical Physics, 31 Caroline St. N., Waterloo, ON, N2L 2Y5, Canada
\\
$^8$ Astrophysics and Cosmology Research Unit, School of Chemistry and Physics, University of
KwaZulu-Natal, Durban, 4001, South Africa
\\
}

\begin{document}
\date{\today}
\pagerange{\pageref{firstpage}--\pageref{lastpage}} 
\pubyear{2015}
\maketitle
\label{firstpage}

\begin{abstract}
Estimating the all-sky rate of fast radio bursts (FRBs) 
has been difficult due to small-number statistics and the 
fact that they are seen by disparate surveys 
in different regions of the sky. In this paper we 
provide limits for the FRB rate at 800 MHz based on the 
only burst detected at frequencies below 1.4 GHz, FRB 110523. 
We discuss the difficulties in rate estimation, particularly in 
providing an all-sky rate above a single fluence threshold. We
find an implied rate between 700-900 MHz that is consistent with
the rate at 1.4 GHz, scaling to 
$6.4^{+29.5}_{-5.0} \times 10^3$\,sky$^{-1}$\,day$^{-1}$ 
for an HTRU-like survey. This is promising for upcoming experiments below 
a GHz like CHIME and UTMOST, for which we forecast detection rates.
Given 110523's discovery
at 32$\sigma$ with nothing weaker detected, down to the 
threshold of 8$\sigma$, we find consistency with 
a Euclidean flux distribution but disfavour steep distributions, 
ruling out $\gamma > 2.2$.


\end{abstract}

\newcommand{\be}{\begin{eqnarray}}
\newcommand{\ee}{\end{eqnarray}}
\newcommand{\beq}{\begin{equation}}
\newcommand{\eeq}{\end{equation}}

\section{Introduction}

A new class of radio transients known as 
fast radio bursts (FRBs) has been 
discovered in the last decade. FRBs are highly dispersed millisecond 
events whose origin remains unknown. Their large dispersion 
measures ($\rm DM \sim360$-$1600$ pc cm$^{-3}$) imply they come 
either from cosmological distances ($z \sim 0.3$-1) or 
regions of over-dense plasma. They occur with 
frequency of thousands per sky per day but the
volumetric rate this implies depends strongly on whether
the large DM observed resides in the intergalactic medium (IGM)
or the host galaxy.
If the dispersion occurs in the IGM the sources are at
cosmological distances and the FRB rate is within a couple orders of magnitude of the 
core-collapse supernova rate for a cosmological population. 

To date, estimating the all-sky rate of FRBs has proven 
difficult, even at 1.4 GHz where most have been found
\citep{2007Sci...318..777L, 2012MNRAS.425L..71K, 
2013Sci...341...53T, 2015MNRAS.447..246P, 2015arXiv151107746C, 2014ApJ...790..101S}. 
This is in part 
because of their unknown flux distribution and location within the radio telescope beam, as 
well as the low number of observed events. It is further 
exacerbated by the different specifications of the surveys that find them, 
whose disparate search algorithms, fluence completenesses, and 
sensitivity can affect their detection rate. Extrapolating to other 
frequencies is also difficult since spectral indices 
and the extent of scattering are still unknown. 

\cite{2013Sci...341...53T} searched about $25\%$ of
the 
High Time Resolution Universe (HTRU) survey data and found four FRBs. 
They estimated 
an all-sky daily rate of $1.0^{+0.6}_{-0.5}\times 10^4$ above 
$\sim3$ Jy ms from 23 days of data and using a $0.55$ deg$^2$
beam. The rate based on HTRU 
has since come down ($6^{+4}_{-3}\times 10^3$ sky$^{-1}$ day$^{-1}$) 
with the discovery of five more FRBs in three times as much 
data \citep{2015arXiv151107746C}. 
\cite{2015MNRAS.447.2852K} also found a rate that was 
lower than the initial estimate, 
calculating $\sim2500$ sky$^{-1}$ day$^{-1}$
after accounting for completeness factors like fluence sensitivity. 
Though the error bars are still significant, there is
some convergence on the rate, and it now seems likely that there are 
thousands of such events each day at 1.4 GHz.

Event rate estimates at 1.4 GHz are converging, 
but strong rate constraints have not yet been made in other bands. 
The non-detection by UTMOST (an upgrade to the Molonglo Observatory Synthesis
Telescope) 
\citep{2016arXiv160102444C} placed 
a 2$\sigma$ upper limit on the number of 
bright events ($10^3$ events per sky per day 
above 11 Jy ms) after 
searching 467 hours at a fraction of its eventual sensitivity. 
FRB 110523 remains the only published event not found around 1.4 GHz. 
It was found near 800 MHz, where scattering or the
intrinsic spectral index might have rendered this lower-frequency FRB 
unobservable. \cite{2015arXiv151109137K} argued that the steep
blue spectrum seen in FRB 121102 \citep{2014ApJ...790..101S} was indicative 
of free-free absorption, the optical depth of which scales as $\lambda^{2.1}$  
and would make metre-wave bursts difficult to see. A greater concern 
comes from scattering. Sources broadened by scattering 
to $\sim10$ ms at 1.4 GHz would 
be $\sim100$ ms at 800 MHz, and a couple of seconds at 400 MHz, 
due to the $\lambda^{4}$ scaling of the scattering width.

Some surveys that could have great
impact on FRB science are threatened by strong scattering.
ALERT hopes to localize dozens 
of bursts with LOFAR after finding them with the large field-of-view 
(FoV) APERTIF \citep{leeu14, 2008AIPC.1035..265V}, 
UTMOST will have $\sim8$ deg$^2$ of sky coverage 24/7 at 843 MHz
\citep{2016arXiv160102444C}, 
and HIRAX\footnote{http://www.acru.ukzn.ac.za/$\sim$hirax/}, 
Tianlai\footnote{http://tianlai.bao.ac.cn/}
and CHIME (400-800 MHz) 
could see 10$^{2-4}$ per year, with the ability 
to write full polarization information \citep{2014SPIE.9145E..22B}. 
However, their success depends on whether or not the 
rate of detectable FRBs is comparable to that at higher frequencies.

\section{Flux distribution}

FRB 110523 was found by searching data from 
the Green Bank Hydrogen Intensity Mapping (GBTIM hereon) 
survey \citep{2010Natur.466..463C, 2015Natur.528..523M, 2013MNRAS.434L..46S}. 
These data were taken with 1.024\,ms cadence between 700-900 MHz 
using the GBT linearly-polarized prime-focus 800\,MHz receiver, along 
with the GBT Ultimate Pulsar Processing Instrument (GUPPI) digital back-end.
An effective DM range of 20-2000\,pc\,cm$^{-3}$ was then searched for FRBs. 
At each DM, the data were convolved with all possible lengths of 
top-hat windows up to 100\,ms to search for peaks. 
The peaks were then compared to the 
root mean square (RMS) of the convolved time-stream, the ratio of 
which is what we will refer to as signal-to-noise ratio (SNR). 
The survey duration was 660 hours.

In order to test the observed FRB flux distribution, $N(F)$, we can apply 
a standard log$(N)$-log$(F)$ test. We will consider only power-law distributions 
of form
$N(>\!F) \propto F^{-\gamma}$.
In a Euclidean Universe a population of sources that are 
uniformly distributed in space 
should have $N(>\!F)\propto F^{-3/2}$. This makes intuitive sense, since 
number counts ought to increase like the cube of distance, while the flux falls 
off as inverse squared distance. 

With no FRBs found between the 
search algorithm's detection threshold, 8$\sigma$, and 32$\sigma$, 
where FRB 110523 was found, we can test if this has any implications for 
the true flux distribution. The question we 
are trying to answer is ``Having seen 
a single event, what is the probability that it has SNR greater than 
$s$ for a given value of $\gamma$?". This is given by the 
ratio of integrals, 
\begin{align}
\label{eq-beta}
\beta \equiv \frac{\int_{s_{\rm max}}^{s_{\infty}} 
N(s) ds}{\int_{s_{\rm min}}^{\infty} N(s) ds},
\end{align}
which reduces to 
$\beta = \left (\! \frac{s_{\rm max}}{s_{\rm min}}\! \right)^{-\gamma}$ for 
$\gamma \neq 0$ and integrands 
of the form $N(s)\propto s^{-\gamma-1}$. This statistic is 
equivalent to the $V/V_{\rm max}$ test that has been used 
to probe the underlying spatial distribution of quasars 
\citep{1968ApJ...151..393S} as well as gamma-ray bursts 
\citep{1991ApJ...383L..61O}. Calculating $\beta$ as a 
function of $\gamma$ shows that steep distributions 
with $\gamma > 2.2$ are ruled 
out with $95\%$ confidence by this single detection alone. 

This is mathematically 
equivalent to the single-burst solution to 
a more general approach similar to the biased 
coin-flip scenario outlined by
\citet{2016MNRAS.tmpL...8C}.
If $M_{\rm high}$ FRBs
are observed above a threshold SNR of $s_{\rm thresh}$, with 
$M_{\rm tot}$ above $s_{\rm min}$, and $p$ is the relative probability 
of detecting an FRB in the high-SNR region, then 
\begin{equation}
\label{eq-binomial}
P(M_{\rm high}| \,  M_{\rm tot}, p) =  \binom{M_{\rm tot}}{M_{\rm high}} 
\, p^{M_{\rm high}} (1-p)^{M_{\rm tot} - M_{\rm high}},
\end{equation}
where $p$ is just $\beta(\gamma)$. Clearly this 
reduces to the previous result in the case where $M_{\rm high}=M_{\rm tot}=1$.

\section{Rates}

The simplest constraints one can make, given a set of 
observations, will be an expected event rate for a future survey with identical 
parameters. Transferring that rate to another survey 
or onto the sky requires care, and in both cases uncertainties 
are introduced that are hard to quantify. For this reason 
we start by calculating a rate for GBTIM in Sect.~\ref{gbtrate}, 
which predicts how many FRBs are expected if an identical 
survey were to take place again. After that we discuss the implications 
for other comparable surveys, which should be fairly robust against 
things like burst-width sensitivity and the choice of 
fluence thresholds. In Sect.~\ref{allskyrate} we provide 
an all-sky rate, with several caveats, and discuss the meaning of 
such a value.

\subsection{Burst rate}

\label{gbtrate}
The rate of FRBs implied by 110523 will be independent 
of its observed brightness. The relevant quantity is the survey
sensitivity, so $s_{\rm min}$ is the only 
flux scale that should show up in our rate estimate. 
For a true rate $\mu_{0}$ above $s_{\rm min}$, we would expect 
the number of bursts, $M$, in a given survey above some SNR, $s$, to be

\begin{equation}
M_s = \mu_{0}\, \Omega 
\, T_{\rm int} \frac{N(\!>\!s)}{N(\! >\!s_{\rm min})},
\end{equation}

\noindent where $\Omega$ is the telescope's FoV and $T_{\rm int}$
is the time on sky. 
We use the rate above some SNR so that we can easily scale it to a rate prediction for a different survey without making any implicit assumption about the distribution of FRBs in flux, fluence, or duration.
Similar to Eq.~\eqref{eq-beta}, this becomes 

\begin{equation}
\label{eq-mu}
M_s = \mu_{0}\, \Omega \, T_{\rm int} 
\frac{s^{-\gamma}}{s_{\rm min}^{-\gamma}} \,\,\,\, \rm{for} \,\, \gamma > 0 .
\end{equation}

However, since we will not try to estimate an all-sky daily 
rate until Sect.~\ref{allskyrate}, for now we will take 
$T_{\rm int}$ and $\Omega$ 
to be in units of the GBTIM on-sky time and beam-size. Therefore 
$M_s$ should be thought of as the number of FRBs one would expect 
if the GBTIM were repeated.

\subsubsection{Frequentist rate limits}

If we regard the sky rate $\mu_{0}$ as fixed, we can immediately 
write down the probability of observing $M_{\rm tot}$ 
FRBs above a SNR of $s$. It is simply given by the Poissonian distribution

\begin{equation}
   P(M_{\rm tot}|\mu_{0}) = \frac{M_{s_{\rm min}}^{M_{\rm tot}} 
   \mathrm{e}^{-M_{s_{\rm min}}}}{M_{\rm tot}!},
\end{equation}

\noindent where $M_{s_{\rm min}}$ is given by Eq.~\eqref{eq-mu} for $s = s_{\rm min}$. 
Now we can ask which values of $M_s$ make the observed value of $M_{\rm tot} = 1$ 
unlikely. Choosing a threshold value of 5\%, 
we can---in this sense---rule out expected event counts $M_{s_{\rm min}}$ 
outside of the range from 0.05-4.50 events per GBTIM-like survey, with 
a maximum likelihood value at 1. 

\subsubsection{Bayesian rate limits}

From a Bayesian viewpoint, we want to look at the 
posterior for the expected number of 
detections, $M_{s_{\rm min}}$ rather than the likelihood. 
For simplicity we choose a flat prior on $M_{s_{\rm min}}$, 
which means that the posterior is again
\begin{equation}
   \label{eq-posterior}
   \mathcal{P}(M_{s_{\rm min}}|M_{\rm tot}) = 
    \frac{M_{s_{\rm min}}^{M_{\rm tot}}
    \mathrm{e}^{-M_{s_{\rm min}}}}{M_{\rm tot}!}.
\end{equation}
Note that, although the posterior has the same functional form as 
the likelihood, it is to be read as a density in $M_{s_{\rm min}}$ 
rather than a probability for $M_{\rm tot}$. 
Now we can calculate another 95\% 
confidence interval, defined as the smallest interval $I$ with the 
property $\int_I \mathrm{d}\mu_{0} \, \mathcal{P}(\mu_{0}|M_{\rm tot}) = 0.95$. 
We find for this 95\% confidence interval
$I = [0.24,\, 5.57]$ events for a GBTIM-like survey. From 
hereon we will quote the rate error bars based on the posterior. 
The posterior for $\mu_0$, which is the same as 
Eq.~\eqref{eq-posterior} multiplied by $\Omega \, T_{\rm int}$, 
is shown in Fig.~\ref{fig-posterior}.

\subsection{Implications for other surveys}
\label{chimerate}

Though only one event was observed in the 660 hours of archival 
data, the fact
that any burst was detectable in this band is significant. 
Some of the most important upcoming surveys for FRB science 
will observe below 1.4 GHz. 
UTMOST \citep{2016arXiv160102444C} will be on the sky 24/7 with an $\sim8$ deg$^2$ FoV 
and 18,000 m$^2$ of collecting area, observing at $843$ MHz. 
ALERT hopes to localize dozens of 
FRBs by first detecting them with the large-FoV APERTIF \citep{leeu14, 2008AIPC.1035..265V}
and
then following up with roughly arcsecond resolution when they 
arrive several minutes later at LOFAR. 
Another survey for which FRB 110523's discovery is relevant
is CHIME, observing at $400$-$800$ MHz. 
If the event rate in this band is comparable to the one at higher 
frequencies, then its large FoV and uninterrupted observing 
will make it by far the fastest FRB survey. 

Since the rate of detection depends on an interplay of the underlying 
FRB flux and scattering distributions 
with a survey's thermal sensitivity, fluence completeness, and observing frequencies,
the comparison of two surveys with similar specifications is by far the safest bet. 
CHIME has $\sim8,000$ m$^2$ of collecting area compared to GBT's $\sim7,850$ 
m$^2$ and has 100 MHz of overlap 
with GBTIM. UTMOST will observe within the GBTIM band with similar 
sensitivity per steradian. 
Though others \citep{2014ApJ...792...19B} have provided models for 
calculating inter-survey sensitivity based on sky pointing 
and temporal broadening, we compare only similar
telescopes and adopt the simplest possible comparison 
based on known features of each instrument. Given how little 
is known about scattering properties and spectral indices, 
we provide only a skeleton model below; a 
more detailed calculation is beyond the scope of this
paper.

A survey, $\Sigma$, that is similar to the GBTIM experiment 
will see $N_\Sigma$ events per day based on the one detected burst
in $\sim27.5$ days at GBT. This is given by

\begin{align}
N_\Sigma =  \frac{1}{27.5} 
\left ( \frac{G_{\Sigma}}{G_{\rm GBT}} 
\frac{\left < T^{\rm sys}_{\rm GBT} \right >}{\left < T^{\rm sys}_{\rm \Sigma} \right >} 
\sqrt{\frac{B_\Sigma}{B_{\rm GBT}}} \right )^\gamma \left 
( \frac{\Omega_\Sigma}{\Omega_{\rm GBT}} \right ) \, \rm{day}^{-1}
\end{align}

\noindent where $B$ gives the survey's bandwidth, $G$ 
is the gain, 
and $\left < T^{\rm sys} \right >$ gives the pointing-averaged
system temperature. For GBT we take the effective solid 
angle based on the full-width half max (FWHM) in power, giving 
$\Omega_{\rm GBT} \sim0.055$ deg$^2$. We use 26.5 K
for the sky-averaged system temperature, and a 
gain of $2$ K Jy$^{-1}$. 

As discussed above, in assessing the impact FRB 110523's detection 
on other surveys, we want to avoid venturing into the unknown. For 
this reason we consider only the 100 MHz of overlap between CHIME 
and GBTIM, since that region is known to have a non-zero rate of 
observable FRBs. For things like beam size, we take the maximum 
possible FoV based on CHIME's optics and let others adjust 
the effective solid angle accordingly; though the CHIME collaboration 
may search only a subset of their primary beam in order to optimize 
other aspects of their FRB survey, we will estimate the rate based 
on a full beam.  

We model CHIME's primary beam at 750 MHz 
based on \citet{2015PhRvD..91h3514S}. 
A simple dipole beam in the aperture plane 
is propagated onto the sky by treating the reflector 
along the cylinder (north-south direction) as a mirror, and 
by solving the Fraunhofer diffraction 
problem in the east-west direction. As with GBT, we use only the 
beam within the half-max contour. This gives 
$\Omega_{\rm CH}\sim86$ deg$^2$ in the middle of 
its band compared to 
$\Omega_{\rm GBT}\sim0.055$ deg$^2$. Though this gives a ratio 
of $\sim1600$ between the two telescope's beam sizes, we remind the 
reader that this is an approximate solid-angle upper-limit 
for CHIME between 700-800 MHz. 
We then estimate its aperture efficiency as 50$\%$, compared with $72\%$ at 
GBT\footnote{https://science.nrao.edu/facilities/gbt/proposing/GBTpg.pdf}, 
whose feed horn maximally illuminates its dish 
while minimizing ground spill, something that is difficult with 
CHIME's dipole antennas.
This makes $G_{\rm CH} = 1.38$ K Jy$^{-1}$.
Finally, keeping with 26.5 K for 
GBT's system temperature as before and using CHIME's design system
temperature of 50K \citep{2014SPIE.9145E..22B}, we can write the 
maximum-likelihood value for the CHIME rate as

\begin{equation}
N_{\rm CH} \approx 7.5 \left ( \frac{50 \,\rm K}{T^{\rm sys}} \right )^{1.5}
\, \rm{day}^{-1}
\end{equation}

\noindent assuming a Euclidean distribution. This means with a  
50 K system temperature, CHIME could see between 2-40 (95$\%)$ 
bursts per day if it searches its whole FoV, based on the known 
non-zero rate above 700 MHz.  
With a more conservative sky-averaged system 
temperature $T^{\rm sys}=100$ K, CHIME might expect between one every couple 
of hours and one every two days. 

\cite{2016arXiv160102444C} estimate the daily rate of UTMOST in a similar way, 
directly comparing their sensitivity with that of Parkes at 1.4 GHz. 
They estimate that they will see a burst once every several days.
However, with our constraints on the rate between 700-900 MHz, we can 
recompute UTMOST's detection rate based on the same band, once 
it reaches final sensitivity. 
We use $G=3.6$ K Jy$^{-1}$, $T^{\rm sys}=70$ K,
$B=31.25$ MHz, and a factor of $1/\sqrt{2}$ for its single polarization, 
based on \citet{2016arXiv160102444C}. 
This gives $4.2^{+19.6}_{-3.2} \times 10^{-1}$ day$^{-1}$, or between a couple per 
day and one every couple of weeks. This is consistent with \citet{2016arXiv160102444C}.

Finally, we estimate rates for three smaller telescopes 
related to CHIME. We use $\gamma=3/2$ and only the 
100 MHz of overlap bandwidth with GBTIM, as before. 
CHIME's Pathfinder, 
which is made of two 20$\times$37 metre cylinders, has been commissioned 
over the last two years and now has a working beamforming backend. 
If the single formed beam were on sky searching for FRBs at all times, 
one might expect to detect $0.4-9$ per year, taking its beam 
to be $\Omega_{\rm PF} = 0.62$\,deg$^2$ and $G_{\rm PF}=0.26$ K Jy$^{-1}$.
The 26 m John A. Galt Telescope, 
just $\sim150$ m from the CHIME Pathfinder 
at the Dominion Radio Astrophysical Observatory (DRAO), could detect 
$0.1-3$ each year, with 
$\Omega_{26} = 0.78$\,deg$^2$ and $G_{26}=0.09$ K Jy$^{-1}$. 
Another telescope to which a simple FRB 
backend could be attached is the 46 m Algonquin Radio Observatory 
(ARO). This might yield $0.2-4.5$ per year, using 
$\Omega_{\rm ARO} = 0.25$\,deg$^2$ and $G_{\rm ARO}=0.29$ K Jy$^{-1}$. Though none of these 
telescopes makes for a very fast survey, the cost of searching 
is quite small, and a coincident detection between DRAO and 
ARO could provide a sub-arcsecond localization.

\subsection{All-sky daily rate}
\label{allskyrate}
The standard method for estimating an all-sky rate given a set 
of observations is to first calculate the rate, $\mu_0$, for that survey 
--- usually the observed number of FRBs divided by the beam size 
and the time on sky --- and then to scale that based on the survey's 
sensitivity threshold and a flux distribution index, $\gamma$. 
This threshold has typically been in fluence, a physically motivated 
quantity for FRBs, and is given by
\begin{equation}
H_{\rm min} = \frac{s_{\rm min} \left < T^{\rm sys} \right >
 \tau}{G \sqrt{m \tau B}},
\end{equation}
where $\left <T^{\rm sys} \right>$ is the pointing-averaged system temperature, 
as before, $s_{\rm min}$ is the SNR threshold used in the search algorithm, 
$G$ is the gain at beam centre, $B$ is the bandwidth, 
$m$ gives the number of polarizations, and $\tau$ is some timescale. 
If one then wants to quote the rate above, say, 3\,Jy\,ms, then the rate 
becomes $\mu \times \left ( \frac{H_{\rm min}}{3\,{\rm Jy}\,{\rm ms}} \right )^\gamma$.

One problem
with this method is that it is not entirely obvious how to choose $\tau$, 
and several groups have approached it differently.
\cite{2015MNRAS.447.2852K} discuss some of these effects and decided to 
use the value at which their survey becomes fluence complete, 2\,Jy\,ms, 
based on the maximum width to which they are sensitive.  
\cite{2016MNRAS.455.2207R} use sampling time, which is the minimum possible 
effective burst width. This will maximise the reported 
search sensitivity because it uses the lowest
possible fluence limit, and therefore generically lowers the final 
rate estimate after scaling to a common fluence. 
A more exact approach 
is to quote the rate above some fluence curve $H \propto \sqrt{\tau}$ between 
$\tau_{\rm min}$ and $\tau_{\rm max}$ corresponding to the actual SNR threshold if white noise is assumed. This is similar to what \cite{2015arXiv151107746C} do, 
who quote their rate above a fluence range.  

\begin{figure}
  \centering
   \includegraphics[trim={0.3in, 0.0in, 0.3in, 0.0in}, width=0.48\textwidth, height=0.36\textwidth]{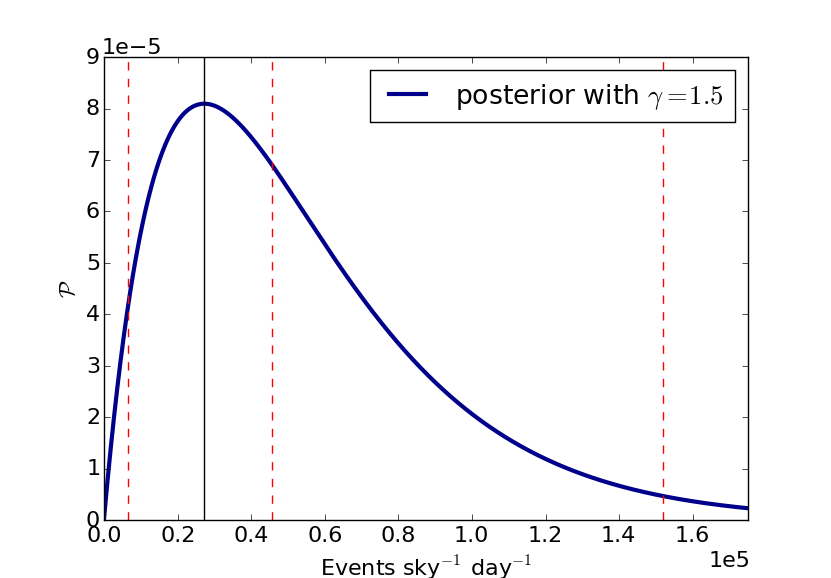}
   \caption{Posterior distribution for the all-sky daily rate 
   based on seeing one burst in 27.5 days of data with a 
   0.055\,deg$^2$ beam. This posterior is meant to be interpreted as the number 
   of FRBs one would see if GBTIM-like surveys were able to
   observe the whole sky for a day, i.e. we have not scaled the rate 
   based on fluence sensitivity
   for reasons described in Sect.~\ref{allskyrate}. 
   The maximum a posteriori value is denoted by the black
   vertical line, which is at 
   $\sim2.7 \times 10^4$\,sky$^{-1}$\,day$^{-1}$. 
   The two outside 
   blue lines enclose 95$\%$ of the curve and the middle blue line 
   denotes the median.}
   \label{fig-posterior}
\end{figure}

Since the primary goal of this paper is to compare between surveys, 
we do not attempt to derive a strict fluence threshold for GBTIM
and to scale our all-sky rate based on it. Until the fluence and 
width distributions for FRBs are known along with 
a search algorithm's width response, the all-sky rate quoted for some incomplete 
region of fluence space is not overly useful. Instead, 
we calculate the rate above our true threshold, 
which is $s_{\rm min}=8$ for 
DMs between 20-2000\,pc\,cm$^{-3}$ and widths between one 
and two hundred milliseconds. A useful estimate of the rate is given by the maximum of Eq.~\eqref{eq-posterior}, 
$\mu_{0} \!=\! \frac{1}{\Omega \, T_{\rm int}}
\left (\! \frac{s}{s_{\rm min}} \! \right )^\gamma$, for 
$s=s_{\rm min}$. The all-sky 
rate for GBTIM above 8$\sigma$ is then 
$2.7^{+12.4}_{-2.1} \times 10^4$\,sky$^{-1}$\,day$^{-1}$, between 700-900\,MHz. 
We plot the corresponding posterior in Fig.~\ref{fig-posterior}.

Though this value seems high,
GBTIM is a sensitive survey, with $F_{\rm min}=97$\,mJy for a 3\,ms 
pulse. Without making any concrete statements about our sensitivity in 
fluence space, we can get an idea of how this rate compares to 
the estimates from other surveys based only 
on thermal sensitivity. We can use the rate 
inferred from the 9 HTRU FRBs as a baseline \citep{2015arXiv151107746C}. 
If we assume the width completeness 
of various surveys is roughly similar, we can tether our rate
to the HTRU one, and calculate a sensitivity ratio, $r_s$. Comparing 
Parkes and GBT, this will be
\begin{equation}
r_s =  
\frac{\left < T^{\rm sys}_{\rm H} \right >}{\left < T^{\rm sys}_{\rm GBT} \right >}
\frac{G_{\rm GBT}}{G_{\rm H}} 
\sqrt{\frac{B_{\rm GBT}}{B_{\rm H}}}
\frac{s^{\rm H}_{\rm min}}{s^{\rm GBT}_{\rm min}}. 
\end{equation}
Using $s^{\rm H}_{\rm min}=10$, an average 
on-axis gain of $G_{\rm H}=0.64$\,K\,Jy$^{-1}$, 
$B_{\rm H}=340$\,MHz, and a 23\,K system temperature \citep{2010MNRAS.409..619K}, 
we find
$r=2.60$. Our rate can then by multiplied by $r^{-\gamma}$,
which gives $6.4^{+29.5}_{-5.0}\times 10^3$\,sky$^{-1}$\,day$^{-1}$, 
assuming a Euclidean distribution. 

This is an extrapolation of our rate estimate at 700-900 MHz
to 1.4 GHz. It corresponds to the number of FRBs that 
HTRU should be detecting if the
intrinsic rates of FRBs in the two frequency bands were the same. This extrapolated 
rate is indeed consistent with the rate observed by HTRU, which 
shows that the rate of FRBs detectable at low frequencies is not 
significantly lower than at 1.4 GHz, which was not previously obvious 
due to the threat of scattering and steep blue power-laws 
\citep{2015arXiv151109137K}. 
This result makes the aforementioned 
upcoming low-frequency surveys especially promising for FRB science. 

This is also consistent with the non-detection upper-limit 
set by \cite{2016arXiv160102444C}, who found the rate to 
be below 10$^3$\,sky$^{-1}$\,day$^{-1}$ for one-millisecond 
11\,Jy bursts at the 2$\sigma$ level. 
This was based on two surveys, one with 467 hours 
on sky, and another with 225 hours on sky at roughly twice the sensitivity. 
Comparing their time-weighted thermal sensitivity with GBTIM, 
we get $r_s\approx10^2$, making our 95$\%$ upper-limit a 
few hundred per sky per day.

\section{Conclusions}
FRB 110523 is the only FRB to be observed below 1.4\,GHz. 
Its detection is encouraging because there are several upcoming 
surveys below a GHz whose impact on FRB science 
is hard to overestimate, so long as the transients are detectable 
at low frequencies. In the next several 
years CHIME, HIRAX, Tianlai, UTMOST, and ALERT could 
increase the number of detected FRBs by orders of magnitude, provide 
polarization information and repetition statistics, and localize them. 
In this paper we have provided the first detailed bounded constraints on the 
FRB rate below 1.4 GHz. 

We have shown two ways of estimating the rate given 
the detection of FRB 110523, 
one based on a frequentist hypothesis test, and the other 
done in a Bayesian framework. These give the same maximum-likelihood 
value, but somewhat different 95$\%$ confidence intervals. We 
have then used the GBTIM estimate to forecast rates for 
CHIME and UTMOST, explicitly only comparing surveys with 
similar specifications. We find CHIME could detect between 2 and 40 per day, given by 
$\approx 7.5 \left ( \frac{50 \, \rm K}{T^{\rm sys}} \right)^{1.5}$ day$^{-1}$, 
making it the fastest upcoming survey. UTMOST, 
which observes in a band inside GBTIM's and whose 
sensitivity per steradian should eventually
be comparable, could see between a couple per day and one 
every two weeks. We also found that CHIME Pathfinder's single 
formed beam, the nearby 26 m John A. Galt Telescope, and the 
46 m ARO might see a couple FRBs each year, providing 
sub-arcsecond localisation through VLBI.

The difficulties of estimating an all-sky rate above a 
single fluence value was discussed. We showed how 
an on-sky rate not attached to a specific survey is 
not only hard to predict but also hard to interpret. For that 
reason we estimated a rate above the true threshold for 
GBTIM --- an SNR of 8 --- which gave us 
$2.7^{+12.4}_{-2.1} \times 10^4$\,sky$^{-1}$\,day$^{-1}$. 
The fluences to which GBTIM was sensitive are those above the 
curve  
$0.17 \sqrt{\left(\tau/\mathrm{ms}\right)}$\,Jy\,ms for pulse widths between 1-100\,ms. To test the agreement between this rate and 
those found by other surveys, we scaled based only on 
thermal sensitivity. If we extrapolate from this daily 
rate to a survey with the sensitivity of HTRU, we find 
$6.4^{+29.5}_{-5.0} \times 10^3$\,sky$^{-1}$\,day$^{-1}$, 
which is consistent with \citep{2015arXiv151107746C}.

We also investigated the flux distribution index, 
$\gamma$, and found that steep distributions with 
$\gamma > 2.2$ are ruled out.

\section{Acknowledgements}

We thank Emily Petroff and Joeri van Leeuwen for useful discussions. 
We also thank NSERC for support. 

\newcommand{\araa}{ARA\&A}   
\newcommand{\afz}{Afz}       
\newcommand{\aj}{AJ}         
\newcommand{\azh}{AZh}       
\newcommand{\aaa}{A\&A}      
\newcommand{\aas}{A\&AS}     
\newcommand{\aar}{A\&AR}     
\newcommand{\apj}{ApJ}       
\newcommand{\apjs}{ApJS}     
\newcommand{\apjl}{ApJ}      
\newcommand{\apss}{Ap\&SS}   
\newcommand{\baas}{BAAS}     
\newcommand{\jaa}{JA\&A}     
\newcommand{\mnras}{MNRAS}   
\newcommand{\nat}{Nat}       
\newcommand{\pasj}{PASJ}     
\newcommand{\pasp}{PASP}     
\newcommand{\paspc}{PASPC}   
\newcommand{\qjras}{QJRAS}   
\newcommand{\sci}{Sci}       
\newcommand{\solphys}{Solar Physics}       %
\newcommand{\sova}{SvA}      
\newcommand{\aap}{A\&A}
\newcommand\jcap{{J. Cosmology Astropart. Phys.}}%
\newcommand{\prd}{Phys. Rev. D}

\bibliography{gbt_rate_mnras}
\bibliographystyle{mn2e}

\label{lastpage}

\end{document}